\documentclass[12pt,preprint]{aastex}
\usepackage[graphicx]{}

\newcommand{\vy}[2]{#1_{\scriptscriptstyle #2}}

\def\gtorder{\mathrel{\raise.3ex\hbox{$>$}\mkern-14mu
             \lower0.6ex\hbox{$\sim$}}}
\def\ltorder{\mathrel{\raise.3ex\hbox{$<$}\mkern-14mu
             \lower0.6ex\hbox{$\sim$}}}
\def\proptwid{\mathrel{\raise.3ex\hbox{$\propto$}\mkern-14mu
             \lower0.6ex\hbox{$\sim$}}}
\def\0946{PG~0946+301}

\def\arcsec{\ifmmode '' \else $''$\fi}

\def\arcsecpoint{\ifmmode ''\!. \else $''\!.$\fi}

\def\kms{\ifmmode {\rm km\ s}^{-1} \else km s$^{-1}$\fi}
\def\Msun{\ifmmode {\rm M}_{\odot} \else M$_{\odot}$\fi}
\def\Lsun{\ifmmode {\rm L}_{\odot} \else L$_{\odot}$\fi}
\def\Zsun{\ifmmode {\rm Z}_{\odot} \else Z$_{\odot}$\fi}

\def\ergscm2{ergs\,s$^{-1}$\,cm$^{-2}$}
\def\icm3{{\rm cm}^{-3}}
\def\icm2{{\rm cm}^{-2}}
\def\qo{\ifmmode q_{\rm o} \else $q_{\rm o}$\fi}
\def\Ho{\ifmmode H_{\rm o} \else $H_{\rm o}$\fi}
\def\ho{\ifmmode h_{\rm o} \else $h_{\rm o}$\fi}

\def\vFWHM{\ifmmode v_{\mbox{\tiny FWHM}} \else
            $v_{\mbox{\tiny FWHM}}$\fi}
\def\CCF{\ifmmode F_{\it CCF} \else $F_{\it CCF}$\fi}
\def\ACF{\ifmmode F_{\it ACF} \else $F_{\it ACF}$\fi}
\def\Halpha{\ifmmode {\rm H}\alpha \else H$\alpha$\fi}
\def\Hbeta{\ifmmode {\rm H}\beta \else H$\beta$\fi}
\def\Hgamma{\ifmmode {\rm H}\gamma \else H$\gamma$\fi}
\def\Hdelta{\ifmmode {\rm H}\delta \else H$\delta$\fi}
\def\Lya{\ifmmode {\rm Ly}\alpha \else Ly$\alpha$\fi}
\def\Lyb{\ifmmode {\rm Ly}\beta \else Ly$\beta$\fi}
\def\Lyg{\ifmmode {\rm Ly}\beta \else Ly$\gamma$\fi}
\def\hi{H\,{\sc i}}

\def\hei{He\,{\sc i}}

\def\ciii{\ifmmode {\rm C}\,{\sc iii} \else C\,{\sc iii}\fi}
\def\civ{\ifmmode {\rm C}\,{\sc iv} \else C\,{\sc iv}\fi}

\def\nv{N\,{\sc v}}

\def\o5007{[O\,{\sc iii}]\,$\lambda5007$}

\def\mgi{Mg\,{\sc i}}
\def\mnii{Mn\,{\sc ii}}
\def\crii{Cr\,{\sc ii}}
\def\Niii{Ni\,{\sc ii}}
\def\mgii{Mg\,{\sc ii}}

\def\siiv{Si\,{\sc iv}}

\def\siII{Si\,{\sc ii}}

\def\caii{Ca\,{\sc ii}}

\def\feii{Fe\,{\sc ii}}

\def\alii{Al\,{\sc ii}}
\def\aliii{Al\,{\sc iii}}

\def\o{\o}
\begin{document}

\title{MEASURING COLUMN DENSITIES IN QUASAR OUTFLOWS: \\
       VLT OBSERVATIONS OF QSO~2359--1241\footnote{Based on observations made with ESO Telescopes at the Paranal Observatories under programme ID 078.B-0433(A)}
}


\author{
 Nahum Arav\altaffilmark{1,2}, 
Maxwell Moe\altaffilmark{1},
 Elisa Costantini\altaffilmark{3},
 Kirk~T.~Korista\altaffilmark{4},
 Chris Benn\altaffilmark{5},
 Sara Ellison\altaffilmark{6}
}

\altaffiltext{1}{CASA, University of Colorado, 389 UCB, Boulder, CO 80309-0389,
I:arav@colorado.edu}
\altaffiltext{2}{Department of Physics, Virginia Tech, Blacksburg, Va 24061; email: arav@vt.edu}
 \altaffiltext{3}{SRON National Institute for Space Research
 Sorbonnelaan 2, 3584 CA Utrecht, The Nether\-lands}
 \altaffiltext{4}{Western Michigan Univ., Dept.\ of Physics, 
 Kalamazoo, MI 49008-5252}
\altaffiltext{5}{Isaac Newton Group, Apartado 321, E-38700 Santa Cruz de La Palma, Spain}
\altaffiltext{6}{Department of Physics and Astronomy, University of Victoria, Victoria, B.C., V8P 1A1, Canada}


\begin{abstract}
We present high resolution spectroscopic VLT observations of the
outflow seen in QSO~2359--1241. These data contain absorption troughs
from five resonance \feii\ lines with a resolution of $\sim$7 \kms\ and
signal-to-noise ratio per resolution element of order 100.  We use
this unprecedented high quality data set to investigate the physical
distribution of the material in front of the source, and by that
determine the column densities of the absorbed troughs.  We find that
the apparent optical depth model gives a very poor fit to the data and
greatly underestimates the column density measurements.  Power-law
distributions and partial covering models give much better fits with
some advantage to power-law models, while both models yield similar
column density estimates.  The better fit of the power-law model
solves a long standing problem plaguing the partial covering model
when applied to large distance scale outflow: How to
obtain a velocity dependent covering factor for an outflow situated at
distances thousands of time greater than the size of the AGN emission
source.  This problem does not affect power-law models.  Therefore,
based on the better fit and plausibility of the physical model, we
conclude that in QSO~2359--1241, the outflow covers the full extent of
the emission source but in a non-homogeneous way.

\end{abstract}


\section{INTRODUCTION}


In recent years, the potential impact of quasar outflows on their
environment has become widely recognized.  (e.g., Blandford \&
Begelman 2004; Scannapieco \& Oh 2004; Vernaleo \& Reynolds 2006).
Observationally, these outflows are detected as absorption troughs in quasar spectra
that are blueshifted with respect to the systemic redshift of their
emission line counterparts.  The absorption troughs are mainly associated
with UV resonance lines of various ionic species (e.g.,
\mgii~$\lambda\lambda$2796.35,2803.53
\civ~$\lambda\lambda$1548.20,1550.77,
\siiv~$\lambda\lambda$1393.75,1402.77
\nv~$\lambda\lambda$1238.82,1242.80).  A large spread in 
maximum  velocity is detected for different quasar outflows: from several
hundred \kms\ to more than 30,000 \kms.  Outflow troughs are known as
intrinsic absorbers to distinguish them from intervening and associated
absorber, which are also detected in quasar spectra (Hamann et al.\
1997; and Barlow 1997).  The lines we study here are intrinsic narrow lines
(see Arav et~al.\ 2002 for full discussion) and are similar to the ones
analyzed by de~Kool et~al.\ (2001).

Reliable measurements of the absorption
column densities in the troughs are crucial for determining almost
every physical aspect of the outflows: the ionization equilibrium and
abundances; number density; distance; and mass flux and kinetic
luminosity.  Column density measurements depend on the absorption
model one adopts for the observed troughs in the spectrum.  

In the case of Interstellar Medium (ISM) and Intergalactic Medium
(IGM) absorbers (the latter also known as intervening absorbers), the
apparent optical depth method is an excellent absorption model.  This method
postulates that the absorber covers the full size of the emission
source homogeneously and that the relationship between the the residual
intensity in the trough ($I$) and the optical depth of the line
($\tau$) is give by $\tau_{ap}\equiv-\ln(I)$. The derived $\tau$ is
then converted to column density using a standard formula (e.g.,
equation 1 in Arav et~al.\ 2001a).

However, our group (Arav 1997; Arav et~al.\ 1999a,  1999b;
de~Kool et~al.\ 2001; Arav et~al.\ 2001b, 2002, 2003; Scott et al.\ (2004)
Gabel et al.\ (2005a)) and others (Barlow 1997, Telfer et~al.\ 1998,
Churchill et~al.\ 1999, Ganguly et~al.\ 1999) have shown that the
apparent optical depth method, is not a good approximation for outflow
troughs (see also \S~3).  Most of this evidence came from data sets
that showed fully resolved troughs from unblended doublets (e.g., the
\nv, \siiv, \civ\ and \mgii\ lines mentioned above).  For all these 
doublets, the oscillator strength of the blue (the shorter wavelength)
transition is  twice that of the red transition.  Therefore,
the expected optical depth ratio is 2:1 in favor of the
blue transition. (It is not exactly 2:1 since the wavelengths differ
by a small amount.)  For the apparent optical depth method to hold, we
must have $\tau_{ap}$(blue)=$2\tau_{ap}$(red) which requires:
$I_B=I_R^2$, where $I_B$ and $I_R$ are the residual intensities of the
blue and red absorption troughs, respectively. Using high
signal-to-noise (S/N) observations of fully resolved and unblended
outflow doublets, it became clear that in most cases $I_B$ is
significantly smaller than $I_R^2$, demonstrating that the
$\tau_{ap}$ model is not applicable for AGN outflows.

To avoid this contradiction 
for cases of unblended doublets, most works use a pure partial covering model
to determine the real optical depth and hence the actual column
density.  This model assumes that only a fraction $C$ of the emission source
is covered by the absorber and then solves for a combination of $C$ and
$\tau$ that will fit the data of both doublet troughs while
maintaining the intrinsic 1:2 optical depth ratio (see \S~3 of Arav
et~al. 2005 for the full formalism, including velocity dependence).
The weakness in this method is that we solve for two unknowns ($C$ and
$\tau$) given two residual intensity equations.  As long as the ratio
of the residual intensities is in the permitted physical range (see
\S~3 of Arav et~al. 2005), such a procedure will always yield a
solution. However, this is also the case for other two-parameter
fitting algorithms, which yield different estimates of column
densities.  A main alternative to the pure partial covering model is
the so called inhomogeneous absorber model (de Kool, Korista \& Arav
2002, hereafter dKKA; see \S~3.2.2 here).  This model successfully matched
the data of two quasar outflows (dKKA), while proving less applicable
for the outflows troughs of Mrk~279 (Arav et~al.\ 2005; but see the
Discussion here).  

Being able to distinguish between these two (and other) absorption
models is important not only for column density measurements, but also
for obtaining a more accurate geometrical picture of the outflow,
including its internal structure.  Since doublet observations are
inherently inconclusive in distinguishing between these methods (or
any other two parameter models), to resolve this issue we need data
sets that show outflow troughs from more than two lines of the same
ion. These troughs must be unblended, fully resolved and have a high
signal-to-noise ratio (S/N).

In this paper we present unprecedented high quality observations of a
quasar outflow, which cover troughs from five \feii\ resonance lines
associated with \feii\ UV multiplets 1, 2 and 3 (see \S~2.3). We use
the strong constraints available from these troughs to test models of
absorption material distribution, and to determine the ionic column
densities in all the troughs seen in this outflow.  The object we
targeted is the extensively studied broad absorption line (BAL)
quasar: QSO 2359--1241 (Arav et~al.\ 2001a; Brotherton et~al.\ 2001).
Having detailed knowledge of the existing Keck/HIRES data, allowed us
to tailor the new observations towards a definitive test for models of
absorption material distribution. In future papers we will use these
measured ionic column densities to determine the ionization
equilibrium of the outflow, it's number density, distance and
ultimately the mass flux and kinetic luminosity associated with the
outflow.

The plan of the paper is as follows: In \S~2 we describe the
observations, data reduction and the phenomenology of the outflow; in
\S~3 we test the goodness of the fit for the five \feii\ resonance
lines obtained from different models of absorption material
distribution; in \S~4 we discuss our results, and in
\S~5 we summarize them.

\section{VLT OBSERVATIONS OF THE QSO 2359--1241 OUTFLOW}

\begin{deluxetable}{lcll}
\tablecaption{\sc VLT/UVES observations of QSO 2359--1241}
\tablewidth{0pt}
\tablehead{
\colhead{Date}
&\colhead{UT Start Time}
&\colhead{Integration}
&\colhead{Mean Airmass}
}
\startdata
	09/30/2006    &   01:05   &    2850 sec     &         1.40 \\
        09/30/2006    &   01:57   &    2850 sec     &         1.19 \\
        09/30/2006    &   02:47   &    2850 sec     &         1.08 \\
	09/30/2006    &   03:38   &    2850 sec     &         1.03 \\
        10/11/2006    &   00:07   &    2850 sec     &         1.49 \\
	10/11/2006    &   00:57   &    2850 sec     &         1.25 \\
	10/11/2006    &   01:46   &    2850 sec     &         1.11 \\                
        10/12/2006    &   00:02   &    2850 sec     &         1.50 \\
\enddata
\label{observations_table}
\end{deluxetable}

\subsection{Data Acquisition and Reduction}

We observed QSO 2359--1241 (radio source NVSS J235953-124148, z =
0.868; Arav et al.\ 2001a) using the Very Large Telescope (VLT)
operated by the European Southern Observatory (ESO) in Fall 2006 (see
table \ref{observations_table} below).  The observations were done in
service mode (i.e., done by the VLT staff at their scheduling, without
active participation of the science team while the observations are
taken) using the UV-Visual Echelle Spectrograph (UVES). All
observations used a slit width of 1.0 arcsec.  In order to cover the
full wavelength range we used two UVES settings. The observations
taken in September 2006 used the following setting: Wavelength centers
Blue=346nm (Grating/Filter=CD1/HER5), Red=580nm
(Grating/Filter=CD3/SHP700) and the observations taken in October 2006
used: wavlength centers B=437nm (Grating/Filter=CD2/HER5), R=860nm
(Grating/Filter=CD4/OG590).  Both settings were read out in 2 $\times$
2 pixel-binned, high-gain mode.  A table of the observations is given
below. The Useful spectral coverage was 3200--9000 \AA\ compared to
4320--7450\AA\ for the Keck/HIRES data.  More importantly, the signal
to noise of the VLT data, over the spectral range covering the five
\feii\ resonance lines, was roughly 3 times higher, where both data
sets have similar spectral resolution.  This large increase in signal
to noise allowed us to preform the analysis we present in this paper,
which was not possible with the Keck/HIRES data.  The extraction of
the spectra was done using the UVES pipeline, which is based on
ECHELLE routines in the data reduction package MIDAS.  A detailed
description of this process can be found in Ballester et al. (2000).

The UVES data reduction pipeline gave excellent data products overall.
However, one issue that produced considerable systematic errors are
spectral undulations that were introduced by the pipeline.  These
undulations mimic narrow emission line structure in the data, with
FWHM of about 10 \AA\ and maximum flux of 20\% above the adjacent
continuum. It is clear that these are introduced by the reduction
pipeline since we have overlapping spectral coverage over part of the
affected range (4795--4950 \AA) and undulations appear only in one of
the settings.  The origins of these emission-line structures probably
come from the bad columns located at the bottom of the lower red CCD
chip (private communication, Cedric Ledoux, UVES instrument
scientist).  Fortunately, the observed absorption troughs in that
spectral region are considerably narrower (about 3 \AA\ full width)
than the FWHM of the undulations. Therefore, we are able to model the
undulations out of the data to a large extent (see \S~2.2).  Careful
examination of the continuum fits in the overlapping spectral regions
lead us to estimate a 2--3\% maximum systematic error associated with
these data reduction undulations. We note that the undulations affect
less than 5\% of the entire spectrum range, but unfortunately this
includes the region of the important \feii\ UV 1 multiplet, which
includes two of the \feii\ resonance lines under study here (2586.65\AA\
and 2600.17\AA).

\subsection{Data co-adding and normalization}

Each of the four exposures for a given setting and wavelength region were
heliocentric corrected to vacuum wavelengths given by the fits header
file.     We then used a single multiplier value for each exposure such
that all spectra had the same mean value.  Rejection of cosmic rays
was done by replacing any flux value 3.5 sigma or more above the mean
of the other 3 values and replacing it by that mean value.  Finally,
the four different spectra were co-added via the weighted average at
each data point based on the given error for the 4 different
exposures. The FWHM of the spectral resolution was determined to
be $8.5\pm0.2$ \kms\ based on the weighted average of fitting 1 -
Gaussian profiles to the normalized flux of several atmospheric
absorption lines.  In view of this resolution, we chose to bin the
data to 7.0 \kms, which slightly over-sample the actual
resolution, before normalization and analysis.

The algorithm used to obtain a model for the unabsorbed continuum
entails initially rejecting expected absorption lines a-priory from
the line lists (see table \ref{atomic_transitions_criteria}).  An
initial cubic fit was applied over every spectral point (pixel)
spanning 125 points on either side of a given spectral point.  Data
points $8\sigma$ or more below the fit value were then rejected as
absorption and the cubic fit was reapplied.  This process of refitting
and rejecting was reiterated at $6\sigma$, $4\sigma$ and then
$3\sigma$.  The chosen continuum at this spectral point was the value
of the last cubic fit for that point. This process was repeated for
all data points in the spectrum.  The resultant continuum fit was
smoothed with a boxcar average of five data points.  To normalize the
data, we then divided the observed flux values with this model
continuum.  For problem areas such as the data reduction undulations,
atmospheric absorption regions, or wide/multiple absorption troughs,
the parameters above were slightly altered to accommodate these areas,
but the basic principle of fitting, rejecting and refitting remained
the same.  The only exception to this normalization process at the
spectral vicinity of the \mgii\ broad emission line (BEL) where a pair 
of $\chi^2$ fitted
Gaussians were used to first divide out the \mgii\ emission before
normalization.

\subsection{Identification of spectral features}

The atomic transitions data used to identify spectral features,
normalize the data (see above) and fit the absorption troughs (see \S~3)
came from Morton (2003) for \feii\ transitions with lower energies ($E_{low}$)
below 1000 cm$^{-1}$ and from Kurucz \& Bell (1995) for all other
transitions. Our assembled line list contains all transitions in the spectral
region of the VLT data for QSO 2359--1241 with constraints $E_{low} < 10000$
cm$^{-1}$ (except for \hi\ and \hei\ that can go up to 160000cm$^{-1}$) and 
the  constraints on their $gf$ values (where $g$ is the degeneracy and 
$f$ is the oscillator strength of the transition) are shown in table 2.

\begin{deluxetable}{lll}
\tablecaption{\sc Transitions included in QSO 2359--1241 line list}
\tablewidth{0pt}
\tablehead{
\colhead{$\log(gf)$}
&\colhead{$E_{low}$ (cm$^{-1}$)}
&\colhead{Ions}
}
\startdata
$>-2.5$ & all &  \hei, \mgii, \alii, \aliii, \siII, \caii\ \\
$> -2.0$ & $E_{low}<1000$  & \feii\ \\
$> -1.5$ & $1000<E_{low}<4000$ & \feii\ \\
$> -1.0$ & $E_{low}>4000$  & \feii\ \\
$> -1.0$ & all & \mgi, \crii,\mnii, \Niii\ \\
$> -0.8$ & all &  \hi\ \\
\enddata
\label{atomic_transitions_criteria}
\end{deluxetable}

In order to identify the spectral features in the spectrum we first
measured the redshift of the main outflow component for the
\feii~$\lambda$2587 resonance line ($z=08594$) and then identified
absorption features of other lines from all ions at that redshift.

Table 3 gives the spectroscopic information for the five \feii\
resonance lines that are the focus of the analysis presented here.  A
table containing all the lines from these multiplet (including those
from excited levels) is given in table 1 of de~Kool et~al.\ (2001).
The oscillator values ($f$) are from Morton (2003).

\begin{deluxetable}{ccc}
\tablecaption{\sc \feii\ resonance lines analyzed in this paper}
\tablewidth{0pt}
\tablehead{
\colhead{multiplet}
&\colhead{$\lambda$ vacuum}
&\colhead{relative $f^a$}

}
\startdata
  1   &  2600.17 &  0.81   \\ 
  1   &  2586.65   &  0.23   \\
  2   &  2382.77 &  1.00   \\
  2   & 2374.46    & 0.10  \\
  3   & 2344.21   &  0.35 \\
\enddata
\label{fe2_resonance_lines} \\
$^a$ - oscillator strength relative to $f(2382.77)=0.32$
\end{deluxetable}

\subsection{Description of outflow absorption features}

\begin{figure}
\epsscale{0.5}
\includegraphics[scale=0.35,angle=90]{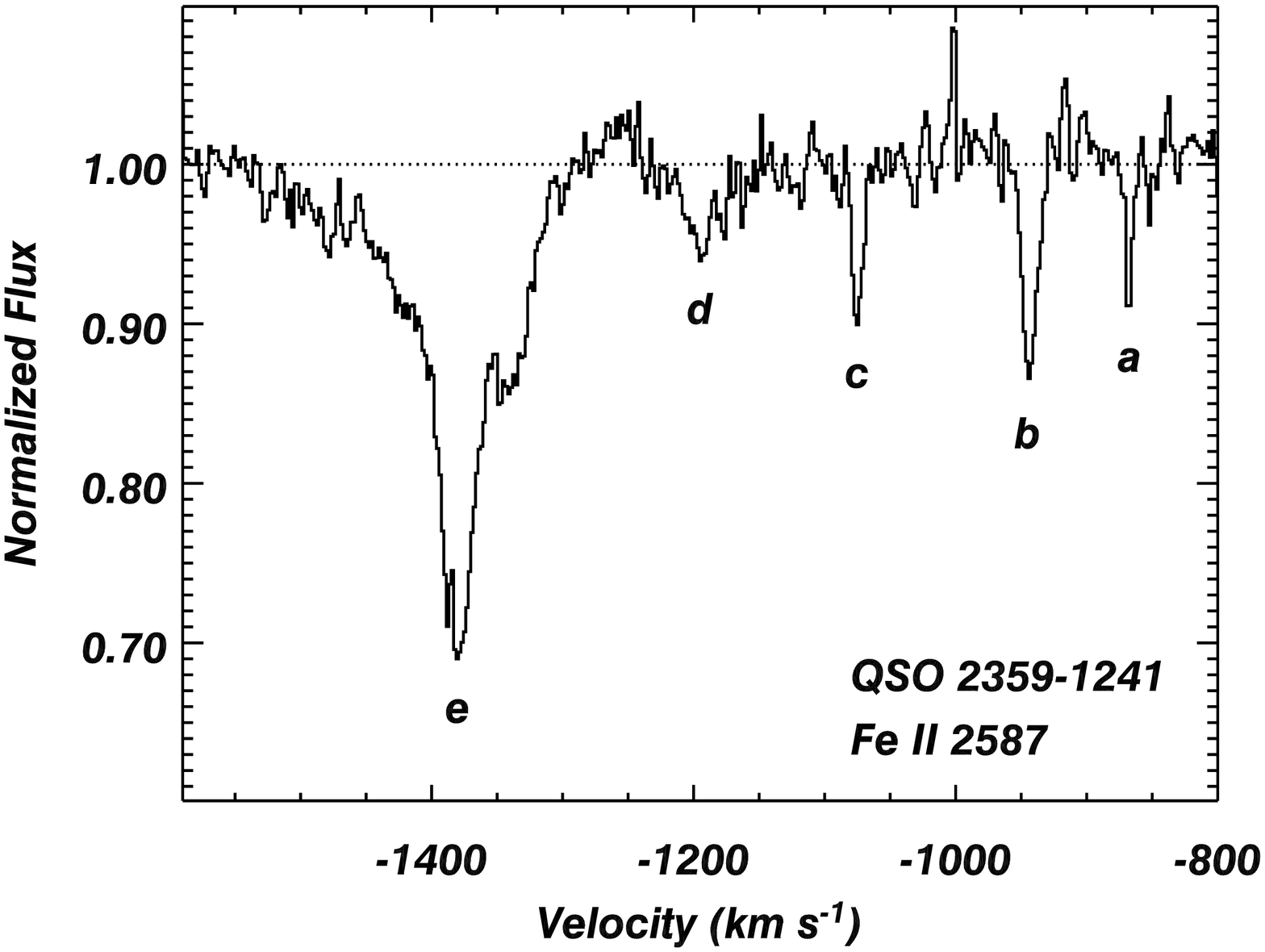}
\caption{Morphology of the outflow in QSO 2359--1241.  Plotted are the
normalized VLT/UVES data for the \feii~$\lambda$2587 resonance
line. The outflow velocity is measured compared to the systemic redshift
($z=0.868$).  The labeling of the components follows that of Arav et~al.\
(2001a),  and most of our analysis is done on component ({\bf e}). }
\label{fig_vel_feii2587.ps}
\end{figure}

In the HIRES paper (Arav et~al.\ 2001a) we have shown that the
absorption features in this object originate from an outflow connected
with the QSO, i.e., intrinsic in nature as opposed to intervening or
associated absorption.  In that paper we also labeled the
outflow components, and since these were unchanged between the two
epochs we will use the same labeling system here.  Figure
\ref{fig_vel_feii2587.ps} shows the absorption components seen in the
\feii~$\lambda$2587 (from the \feii\ UV 1 multiplet) resonance line.
Most of our analysis in this paper will concentrate on the widest and
highest velocity  component, namely {\bf e}.  This component is seen in all the
outflow absorption features detected in this object.  In figure 2 we
show the normalized flux for the spectral range of the \feii\ UV 1
multiplet, where component {\bf e} is labeled for all the lines from
this multiplet.  Most of the troughs arise from low excitation levels
associated with the \feii\ UV 1 multiple.  The on-line version of this
figure show the entire spectrum (spread over 46 pages) where the
absorption features of all the detected ions are identified in a
similar way.

\begin{figure}
\epsscale{0.5}
\includegraphics[scale=0.65,angle=90]{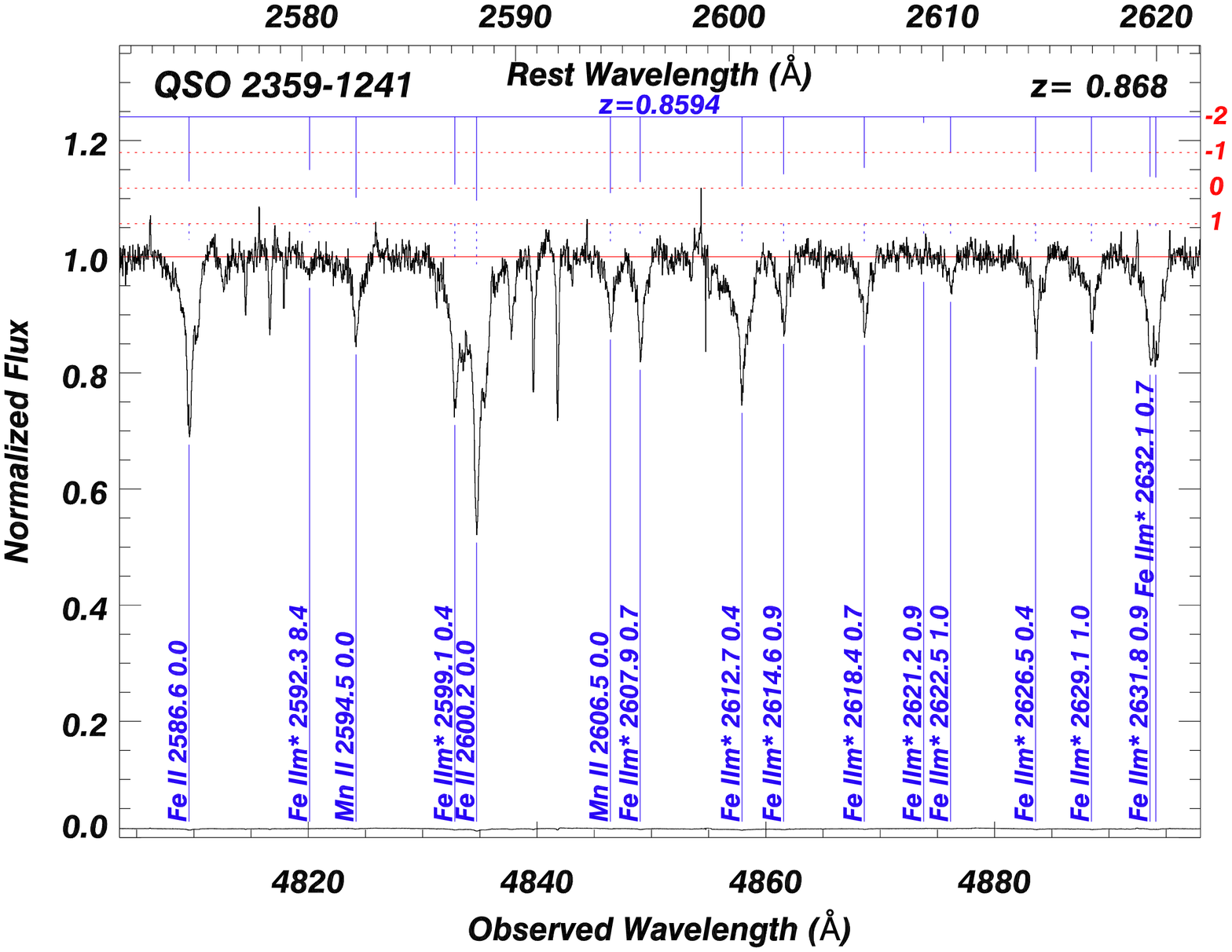}
\caption{ A small portion of the normalized VLT/UVES spectrum of QSO
2359--1241, showing the outflow features associated with the \feii\ UV
1 multiplet.  (A similar presentation of the entire spectrum is
available on-line).  The data plotted on
Fig. \ref{fig_vel_feii2587.ps} are shown here between 4806--4819 \AA\
(observed). Blue vertical lines denote the expected position of
component {\bf e} for various transitions, based on the $z=0.8594$
redshift of this component. Most of these transitions are associated
with \feii\ UV 1 multiplet (two transitions from \mnii\ and one from a
higher \feii\ multiplet are also marked). The size of the
identification line (measured from the top solid line) gives the
$\log(gf)$ value of the transition (where $f$ is the oscillator
strength and $g$ is the degeneracy of the transition), which can be
read from the values associated with the dotted lines (right axis).
Each transition is labeled by its ion, wavelength and energy level
(cm$^{-1}/1000$).  The three unlabeled troughs around 4840 \AA\ and
similarly around 4815 \AA, are components {\bf b,c and d} of
\feii~$\lambda$2600 and \feii~$\lambda$2587, respectively. The error
spectrum is also plotted (just above the X-axis).
The short dotted vertical lines just above the continuum  
show the exact position of the trough around the continuum level.
}
\label{fig_id_feii.ps}
\end{figure} 


\section{MODELING THE ABSORPTION TROUGHS}

As stated in \S~1, in order to extract quantitative information about
the outflow, we need to measure the column density associated with the
absorption troughs.  In this section we test four models for the 
distribution of absorption material in front of the
emission source, in order to find which one gives the best fit for the 
VLT data of QSO 2359--1214.

\subsection{Failure of the Apparent Optical Depth Model}

The apparent optical depth method implicitly assume that the absorption
material covers the source completely and homogeneously.  That is, all
light rays that arrive at the observer pass through material with the
same optical depth at a given wavelength.  Under these assumptions,
the residual intensity of a given line is $I_1=e^{-\tau_{ap1}}$. Under
these assumptions, the optical depth ratio for two E=0 lines from the
same multiplet is $R_{21}=(f_2\lambda_2)/(f_1\lambda_1)$, where $f$ is
the oscillator strength and $\lambda$ is the wavelength of the
transition.  Combining the two relations we derive the expected
residual intensity of line 2 given the observed residual intensity of
line 1: $I_2=e^{-R_{21}\tau_{ap1}}=I_1^{R_{21}}$

\begin{figure}
\epsscale{0.5}
\plotone{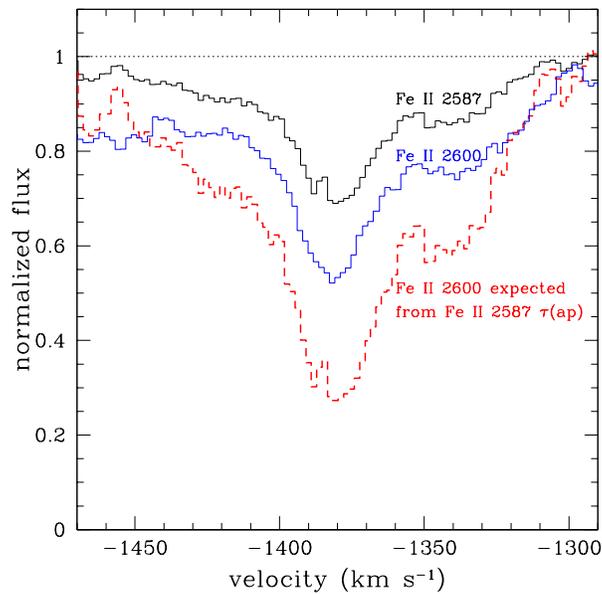}
\caption{Failure of the apparent optical depth model ($\tau_{ap}$).
Normalized residual intensity for \feii~2587 \AA\ and \feii~2600 \AA\
are shown for the main component of the outflow (solid histograms).
The dashed red histogram shows the expected depth of the \feii~2600
\AA\ based on the residual intensity of the \feii~2587\AA\ line and
using the $\tau_{ap}$ model (see text for more details).  It is
evident that the expected depth is a poor fit to the observed
\feii~2600 trough, demonstrating the inadequacy of the $\tau_{ap}$
assumption for this outflow. Similar results are obtained for the
majority of quasars outflows.}
\label{fig_fe2_ap_tau}
\end{figure}

In figure \ref{fig_fe2_ap_tau} we compare the residual intensities of the
outflow troughs arising from two resonance lines ($\lambda2587$ and
$\lambda2600$) of the UV 1 \feii\ multiplet, with the apparent optical
prediction for $I(\lambda2600)$ based on $I(\lambda2587$), using the
above relationship ($R_{21}=3.5$ for this case). The inadequacy of the
$\tau_{ap}$ model in predicting $I(\lambda2600)$ is clear,
demonstrating that the $\tau_{ap}$ model assumptions are invalid for
the \feii\ resonance troughs.  Similar results are obtained for other
ions.  We therefore, need to explore more complicated models.  Below,
 we describe three such models: partial covering,
inhomogeneous absorber and a modified partial covering.

\clearpage

\subsection{Two And Three Parameter Models}

\subsubsection{partial covering model}

\begin{figure}
\epsscale{0.5}
\includegraphics[scale=0.42,angle=-90]{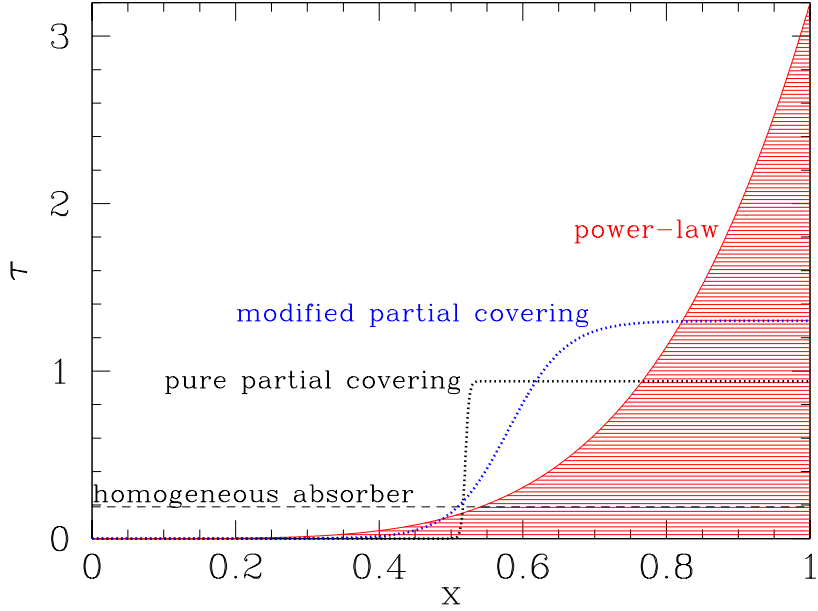}
\caption{The four distributions of absorbing material that
 we investigate in this paper.  
The constant emission source coincides with
the $x$ axis, and the observer is far above the plot.  For the
power-law case we shaded the region of absorbing material, and for all
distributions the absorbing material is below  their curves. For each
distribution the parameters are representative of the deepest part 
in the {\bf e} component of the outflow, where 
a $\chi^2$ fit for all 
five resonance lines was performed at that velocity, and the best fit model
for each functional form is  shown,  $\tau$ is that of the 
 \feii~2587\AA\ line. For the different distributions, the 
optical depth in front of an $x$ location is given
by: 
\newline a) Homogeneous absorber (apparent optical depth case): $\tau(x)=0.19$
\newline b)  partial covering:
$\tau(x)=0$ for $0<x<\vy{x}{0}$, and $\tau(x)=\tau_{\rm{max}}$ for  
$\vy{x}{0}<x<1$; $\tau_{\rm{max}}=0.94, \ \ \vy{x}{0}=0.52$
\newline c) Power-law: $\tau(x)=\tau_{\rm{max}}x^a$, \ \
$\tau_{\rm{max}}=3.2, \ \ a=4.6$ 
\newline d) Modified partial covering: $\tau=\vy{\tau}{0}(1-1/[e^{b(x-\vy{x}{0})}])$, \ \ $\vy{\tau}{0}=1.3$, \ \ $b=25$,  \ \  $\vy{x}{0}=0.58$
\newline
}
\label{optical_depth_distribution_paper_figure.ps}
\end{figure}

Applying the apparent optical depth method is in essence using a one
parameter model ($\tau_{ap}$) for fitting the absorption troughs
($\tau_{ap}\equiv-\ln(I)$).  Two parameter models are the obvious next
step.  Most of the previous work, which addressed the inadequacy of the apparent
optical depth method in modeling outflow troughs, dealt with data of
unblended doublets (e.g., Barlow 1997; Arav et~al.\ 1999a; Ganguly
et~al.\ 1999 Arav et~al.\ 2002; Scott et al.\ 2004). The preferred
method for modeling these troughs is the partial covering model, which
is a two parameter model (Hamann et al.\ 1997; and Barlow 1997). To
summarize this method: we assume that only a fraction $C$ of the
emission source is covered by the absorber and then solve for a
combination of $C$ and optical depth ($\tau$) that will fit the data
of both doublet troughs while maintaining the intrinsic 2:1 optical
depth ratio (see \S~3 of Arav et~al. 2005 for the full formalism,
including velocity dependence).  Implicit in this model are the
assumptions that the absorber covering the fraction $C$  of the source
has the same optical depth across this area, and that the rest of the
source is covered by material with $\tau=0$ in that transition.  A
geometrical illustration of this assumption is shown in figure
\ref{optical_depth_distribution_paper_figure.ps}. The partial covering
model allows us to find a good fit for the doublet troughs and under
the model assumption, a consistent determination of the optical depth
and hence the column density of the trough. This is an important
improvement over the apparent optical depth method, which cannot yield
a consistent optical depth estimates for the two troughs.

As stated in \S~1, the weakness of the partial covering model as
applied to doublet troughs is that we solve for two unknowns ($C$ and
$\tau$) given the two residual intensity equations of the doublet
troughs.  As long as the ratio of the residual intensities is in the
permitted physical range (see \S~3 of Arav et~al. 2005), such a
procedure will always yield a solution.  Therefore, a good fit is a
necessary but not sufficient condition to validate the underlying
model.  In addition, other two-parameter models may fit the data
equally well, where the most studied alternative are inhomogeneous
absorption models

\subsubsection{Inhomogeneous absorption models}

A more general way to modify the apparent optical depth assumptions,
is to assume that the absorption material does not cover the source
homogeneously.  Since there is an infinite number of ways of doing so, we
need a quantitative model that can be directly compared with the
partial covering results.  The first attempt to study generic
inhomogeneous distributions of absorbing material in AGN outflows was
done by de Kool, Korista \& Arav (2002, dKKA).  A formalism
was developed to simulate the effects of an inhomogeneous absorber on
the emerging spectrum and several examples of fitting existing data
with this model were presented.  Arav et al (2005) built upon this
formalism and produced a two parameter inhomogeneous model that can be
directly compared with the partial covering results.  For full details
we refer the reader to these two papers and here we give a brief
description of the idea based on \S~2 of Arav et al (2005).
Our starting point is equation (6) from dKKA:

\begin{equation} 
F(\lambda)=\int\int S(x,y,\lambda)e^{-\tau(x,y,\lambda)}dxdy,
\label{eq:general}
\end{equation}
where $S(x,y,\lambda)$ is the surface brightness distribution of the
background source and $\tau(x,y,\lambda)$ is the line of sight optical
depth at wavelength $\lambda$ in front of a specific $(x,y)$ location,
as defined by dKKA equations (3-5).  

In its most general form, Equation (\ref{eq:general}) is not practical
for modeling spectra.  Both $S(x,y,\lambda)$ and $\tau(x,y,\lambda)$
are unconstrained two variable functions for each  given $\lambda$.  We
need a  simpler and testable model.  To this end, we introduce the
following simplifying assumptions (whose physical
validity and plausibility are discussed in Arav et~al.\ 2005).

\begin{enumerate}
\item $S(x,y,\lambda)=S(\lambda)$, which can be set to $S(\lambda)=1$
when dealing with normalized data.

\item $\tau(x,y,\lambda)=\tau(x,\lambda)$, this simplifying assumption
entails little loss of generality.  We can think of
$e^{-\tau(x,\lambda)}$ as the integrated attenuated flux along a $y$
strip at a specific $(x,\lambda)$.

\item $\tau(x,\lambda)=\tau_{\rm{max}}(\lambda)x^a$, which yields a
two parameter model that can be directly compared with the partial
covering model. We note that a Gaussian characterization also yields a
two parameter model. However, these models were investigated by Arav
et al.\ (2005; see their discussion of figure 3) and were shown to
yield very similar results to power-law models. We therefore use 
the power law model as the sole representative of 2 parameter 
inhomogeneous absorption models.

\end{enumerate}

With the last simplification we have a power-law inhomogeneous model
where each of the two parameters ($\tau_{\rm{max}}$ and $a$) can be
velocity dependent, similar to $C$ and $\tau$ for the partial covering
model.  In general, these two models have quite different physical
interpretations, as will be discussed in \S~4.

\subsubsection{Modified partial covering model}

The pure partial covering model is unphysical since there is a sharp
edge (essentially a step-function) between a region with finite
optical depth and a region with $\tau=0$.  Physically, there need to
be a transition zone between these two regions.  This motivated us to
add a parameter to the partial covering model which controls the
gradient between these two regions.  The phenomenological formalism we chose
is 
\begin{equation}
\tau=\vy{\tau}{0}(1-1/[e^{b(x-\vy{x}{0})}]), 
\label{eq:modified}
\end{equation}
where $\vy{x}{0}$ and 
$\vy{\tau}{0}$ are similar to $C$ and $\tau$ for the pure 
partial covering model and  $b$ control the gradient.

In figure 4 we show the optical depth distribution across the source
for all four models.  The examples we show are tailored to the best
fits of each model to the \feii\ $E=0$ lines (see \S~3.3)

\subsubsection{Conclusive testing of the absorption models}

In order to test the partial covering and the inhomogeneous absorption
models we need data sets that will over-constrain them, that is fitting
more than two equations with the same two unknowns (or three in the
case of the modified partial covering model). To do so we need to
observe more than two troughs from the same ion.

Our VLT/UVES observations of QSO 2359--1241 yield the best such
data-set to date.  In the outflow of this object we detect absorption
troughs associated with more than 30 lines of the \feii\ multiplets UV
1, UV 2 and UV 3 (see figure 2 for the spectral region of the \feii\
UV 1 multiplet).  These lines include five transitions from the $E=0$
energy level, which in principle can give us five residual intensity
equations to fit the two and three free parameters of the above
models.  To realize this possibility, we chose this target, S/N and
spectral resolution, such that we would acquire an unprecedented data set
for these purposes.  The troughs are fully resolved, have
S/N$\gtorder80$ per resolution element and show only minor blending
with other troughs, compared to other objects showing intrinsic 
\feii* troughs (e.g., Wampler, Chugai \& Petitjean 1995; de~Kool et~al.\ 2001, 2002).

\clearpage

\subsection{$\chi^2$ fitting}

In order to get a quantitative measure for how well each model fit the
data, we use $\chi^2$ minimization.  Since the trough is well
resolved, we binned the data to match the measured resolution.  The
resultant data product in the case of the \feii\ resonance lines are
residual intensities for five $E=0$ lines covering the span $-1475$ to
$-1305$ \kms.  Each velocity bin is treated as an independent
measurement and we use a least-squares $\chi^2$ minimization routine
to find the best parameters of each model for that specific bin.  The
process is repeated for all velocity bins.  Two of the troughs suffer
a moderate amount of blending with other lines and we omit the
contaminated region of the individual trough while performing the fit.
These contaminations occur for the \feii~2600\AA\ trough between
$-1475$ to $-1420$ \kms\ and for the \feii~2344\AA\ trough between
$-1340$ to $-1305$ \kms. As a result, we always fit at least four data
points with each model, and five for the $-1420$ to $-1340$, which
includes the deepest part of the trough.  Since the maximum number of
parameters we fit is three, the models are always over-constrained.

To compare the goodness of the fit for each model we calculate the reduced 
$\chi^2$ defined in the following way (see Press et~al.\ 1989 , chapter 14):
\begin{equation} 
\chi^2_{\rm red}=\frac{\sum_{i,j} \left( [I_{i,j}-M_{i,j}]/\sigma_{i,j} \right)^2}{N_{tot}-N_vN_p},
\label{eq:red_chi2}
\end{equation}
where sum $i$ is taken over the number of fitted data point in a given
velocity bin (four or five in our case) and sum $j$ is taken over the
25 velocity bins we fit;  $I_{i,j}$ is the residual intensity of
trough $i$ at velocity bin $j$; $M_{i,j}$ is the modeled residual
intensity of trough $i$ at velocity bin $j$; $\sigma_{i,j}$ is the
associated error for each data point; $N_{tot}$ is the total number of
velocity bins of every transition used in the fit; $N_v$ is the number
of velocity bins (25 in this case) and $N_p$ is the number of free
parameters for each velocity bin (1 for the apparent $\tau$ model, 2
for the partial covering and power-law models, and 3 for the modified
partial covering model). We note that since the minimization procedure 
 omits the blended regions of two troughs, $N_{tot}=112$ and not 125.

\begin{figure}
\includegraphics[scale=0.7,angle=90]{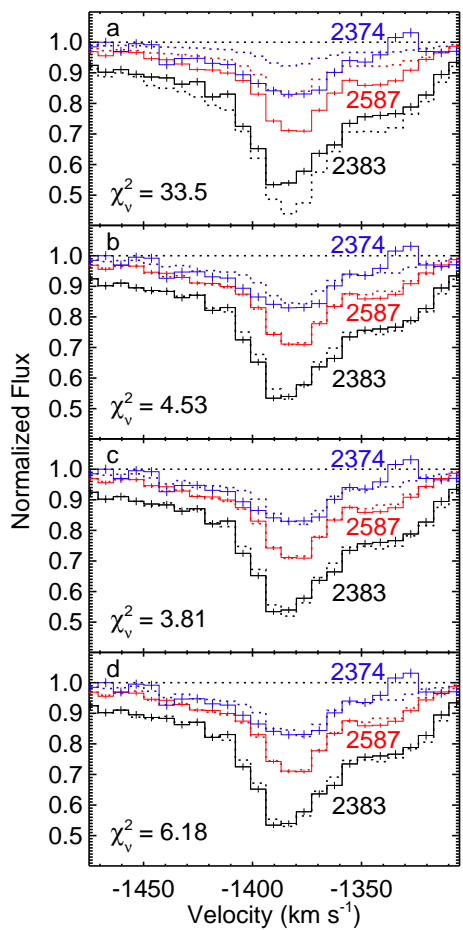}
\caption{Best $\chi^2$ fits for the  \feii\ $E=0$ lines seen in the spectrum of 
QSO 2359--1241. Each panel shows the fit for a different model:
{\bf a} $\tau_{ap}$ model; {\bf b} pure partial covering; {\bf c} power-law model; 
{\bf d} modified partial covering (see text for more details). The data plus
error bars are shown in solid histograms, the fits are shown in dotted histogram,
and the wavelength (\AA) of each transition  is labeled. Best fit reduced $\chi^2$ values are printed at the lower left corner for each model.}
\label{fig_fe2_fit.eps}
\end{figure} 

\clearpage

Figure 5 shows the best $\chi^2$ fit for each model, where for the
sake of clarity we only show three of the five \feii\ $E=0$ lines
detected in the spectrum.  The fits were done using all five lines and
those that are shown span the full range in oscillator strength.
From both visual inspection and from the $\chi^2_{\rm red}$ value, it
is clear that the $\tau_{ap}$ model gives a poor fit to the data.
This is simply a more quantitative way to arrive at the conclusion shown in
figure \ref{fig_fe2_ap_tau}.  The important results shown in figure 5, are:
\begin{enumerate}
\item Both the pure partial covering and the power-law distribution models
give similar fits (with a small advantage to the power-law model). 
\item The modified partial covering model does not yield an improvement 
in $\chi^2_{\rm red}$ value, although the fit itself is somewhat better.
This surprising result will be discussed in \S~3.4.
\end{enumerate}

Before we compare the various fits and discuss how physical they are,
we need to address the absolute $\chi^2_{\rm red}$ values. Normally we
would expect the $\chi^2_{\rm red}$ value of a ``good'' fit to be
$\sim1$.  For our leading two models we find $\chi^2_{\rm
red}$(PL)=3.8, $\chi^2_{\rm red}$(PC)=4.5 (where PL stands for
power-law and PC for pure partial covering).  Nominally, these values
exclude these model at a very high probability. For a model with 60
degrees of freedom a $\chi^2_{\rm red}=3.8$ has less than $10^{-10}$
probability of being the correct model.  However, the main issue here
is the systematic errors that are not addressed by the fit.  First, as
was noted in \S~2, the undulations superimposed on the spectrum due to
data reduction issues can contribute a 2--3\% systematic error in the
normalized intensity of the troughs. The troughs that are most
affected by these undulations (2600\AA\ and 2587\AA) also have the
highest S/N value per resolution element (upwards of 80). As a result
this systematic error by itself can produce a $\chi^2_{\rm red}\sim4$,
even if the model matches the data perfectly.  Another important
source of systematic error is caused by uncertainties in the
oscillator values of the \feii\ resonance lines.  Relative
uncertainties between these values can produce a considerable
$\chi^2_{\rm red}$ on its own.  We tested this hypothesis by
artificially changing the oscillator strength for the weakest
transition (2374 \AA) and found considerable changes in the
$\chi^2_{\rm red}$ values. for example, for the PL model, the value
dropped from 3.8 to 3.4 when the oscillator strength of \feii~2374~\AA
was increased by 40\%.

We therefore are not troubled by the fact that in this actual case we
obtain $\chi^2_{\rm red}\sim4$ and we focus our discussion on the
differences seen between the different models.  It will be left to
future data sets, with better control of the systematics to show
whether these models give reasonable absolute  values of
$\chi^2_{\rm red}$.

\subsection{Comparison between the different models}
 
As noted above, it became clear over the past 10 years that the
simplest possible model ($\tau_{ap}$) is inadequate for modeling
outflow troughs.  Our analysis here only adds a well constrained
quantitative case to this effect.  The next step is to compare between
the two parameter models, namely the pure partial covering and the
power law models.  These models represent rather different physical
pictures with regards to the distribution of absorbing material in
front of the emission source.  We will discuss their pros and cons in
\S~4, but as a first step it is useful to make a simple comparison of
the fits to the same data by models with the same number of free
parameters.

We start by assuming that the main reason for the high absolute values
of $\chi^2_{\rm red}$ for these models are the systematic errors
discussed above.  By that we accept both models as being reasonable
good fit for the data.  The $\chi^2_{\rm red}=3.8$
of the power law model is better than the $\chi^2_{\rm
red}=4.5$ of the pure partial covering model.  However, assessing the
significant of this difference in lieu of the large contribution from
systematic error is more problematic, since it is difficult to predict
what will happen to the relative differences between the $\chi^2_{\rm
red}$ once the systematics are better controlled.  For the mean time
we will cautiously accept the better fit of the power law model.  As we
will see below evidence from other \feii\ transitions and from other
ions lends support to the better applicability of the power law model
as well.  

It is also important to compare the velocity dependence of these
models.  For the partial covering model we find the known result that
the covering factor traces the shape of the strongest modeled
resonance line $C(v)\approx 1-I(v)$.  This behavior is seen in most
covering factor analysis of AGN outflows (e.g., Arav et al.\ 1999b;
Scott et al.\ 2004).  For the power-law model we find that the lowest
exponent is achieved at the point of lowest residual intensity. We
also find that for velocities redwards of the lowest residual
intensity the value of the exponent is modest $4<a<9$, with errors of
$\sim20$\%.  At higher velocity the value of $a$ climbs to roughly 20
with 50\% error typical (1$\sigma$ errors based on the fitting
process).  For such high exponent values the absorption behavior of
the power-law model becomes increasingly similar to that of the pure
covering factor model. The technical reason for the high value of $a$
between $-1450< v < -1410$ \kms\ as compared to the interval $-1370< v
< -1320$ can be seen in figure \ref{fig_fe2_fit.eps}. The intensity
ratio of the strong to weak lines is considerably smaller at the
$-1450< v < -1410$ \kms\ interval compared to the $-1370< v < -1320$
interval.  Independent of a specific model, this behavior suggests a
real difference in the physical behavior of the material in the two
velocity intervals.  Under the assumption of the power-law model, a
large exponent value can be interpreted as material spread in smaller
concentrated entities, but at this stage such an interpretation is
tentative at best.

The modified covering factor model is an attempt to address the
inherent unphysical nature of each of the above models. In the pure
partial covering model there is an unphysically sharp edge
(essentially a step-function) between a region with finite optical
depth and a region with $\tau=0$. Physically, there should be a
transition zone between these two regions. The modified covering
factor model solves this issue by adding a parameter to the partial
covering model which controls the gradient between these two
regions. For the power law model, the entire region where $\tau>3$
yields virtually no flux. Since the data require steep power laws with
high values of $\tau_{\rm max}$, in some cases most of the material
comes from the $\tau>3$ region (see discussion in
Arav et al. 2005). It is disconcerting to have most of the material
coming from regions that do not influence the spectrum much. 
The modified covering factor model allows for a gradual distribution
of material, but caps $\tau_{\rm max}$ at a physically reasonable
value.  However, for the QSO~2359--1241 data set $\tau_{\rm max}$, the
weakest \feii\ resonance line ($\lambda2374$), has $\tau_{\rm max}<3$
across the entire velocity range, except for two data points, where
$\tau_{\rm max}<3.5$.  Therefore, there is no strong saturation at any
point in front of the absorber (not even at $X=1$ where
$\tau=\tau_{\rm max}$).

This maybe the reason why the modified covering factor model gives a
worse fit for the data ($\chi^2_{\rm red}=6.18$).  That is, adding
another degree of freedom was not necessary since the data could be
reasonably fitted with a non saturated power-law. Alternatively, if most of
the $\chi^2_{\rm red}$ comes from systematic errors (the normalization
of the effective continuum and/or oscillator strength issues discussed
in \S~3.2) we will expect a higher $\chi^2_{\rm red}$ when the numbers
of free parameters is increased.  This is because while the $\chi^2$
does not change much, the number of degrees of freedom drops
considerably.  In this actual case, the denominator of equation
(\ref{eq:red_chi2}) changes from 65 to 40, while moving from a 2 to 3
parameter fit, predicting an increase of 62\% in $\chi^2_{\rm red}$
for the same $\chi^2$ value.  This is almost exactly the excess in
$\chi^2_{\rm red}$ for this fit compared with that of the power-law
model. It is therefore possible that the statistical fit of this of
the modified covering factor model will improve enough to make it a
good candidate for modeling the absorption material distribution, once
the systematic errors are under better control.

\subsection{Differences in Inferred Column Densities}

\begin{figure}
\includegraphics[scale=0.37,angle=90]{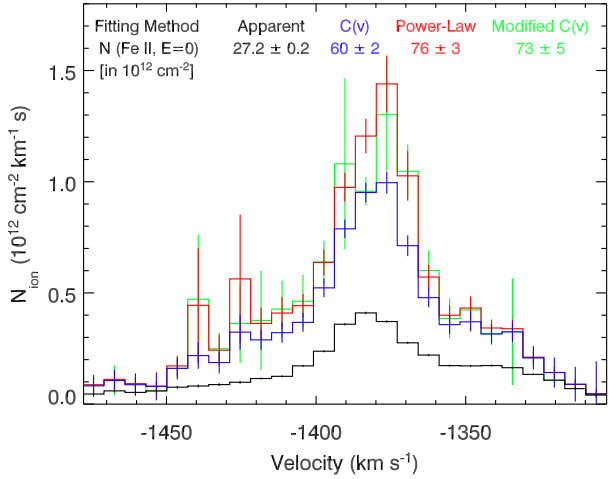}
\caption{Column density of the  \feii\ E=0 level for component {\bf e} of the outflow.
Shown are the best fit determination using the four models discussed in the text.
The top table gives the integrated column density for each of these methods
and identifies each of the curves with the corresponding model. 
 }
\label{fig_fe2_col.eps}
\end{figure}

Figure \ref{fig_fe2_col.eps} shows the extracted column density of the
\feii\ E=0 level for component {\bf e} of the outflow
($N_{\rm{\feii(E=0)}}$), for all four methods. The column densities shown
are the average value across the emission source at each velocity bin.
For example, for the pure partial covering, we multiply the extracted
column density by the covering factor ($C(v)$), since the model
assumes that only a $C(v)$ of the emission source is covered by the
absorbing material.

Several important conclusions can be drawn from figure
\ref{fig_fe2_col.eps}.  First, the column density derived from the
$\tau_{ap}$ model is roughly a factor of 3 lower than those derived by
the other methods.  Not only the $\tau_{ap}$ model inapplicable to AGN
outflow troughs, it also severely underestimates the column density.
Second, the inferred column density for the pure covering factor and
power-law methods differ by only 25\%.  Although these models may
suggest a rather different physical picture of the absorbing material,
this quantitative similarity shows that the ionization equilibrium
results based on their modeled column density will be quite similar.
Furthermore, this last conclusion is strengthened by observing that
not only is the total derived column density similar, but so is the
distribution over velocity.

\subsection{Results for excited \feii\ and \hei}

\begin{deluxetable}{lcll}
\tablecaption{{\sc $\chi^2_{\rm red}$ fits for excited} \feii\ {\sc and} \hei\ {\sc levels}}
\tablewidth{0pt}
\tablehead{
\colhead{energy level$^a$}
&\colhead{lines$^b$}
&\colhead{ $\chi^2$({\small PC})}
&\colhead{ $\chi^2$({\small PL})}
}
\startdata
\feii(0)    &  5   & 4.5   &  3.8   \\ 
\feii(385)    &  7   & 5.8   & 4.8    \\
\feii(668)    &  8   & 5.5   &  4.9   \\
\feii(863)    &  6   & 3.0   &  2.9   \\
\feii(977)    &  5   & 1.4   &  1.3   \\
\feii(7955)    & 4    & 2.4   &  2.3   \\
\hei(159856)    &   5  & 5.0   & 4.3    \\
\enddata
\label{excited_level_chi2}
$^a$ - in cm$^{-1}$\\
$^b$ - number of lines used in the fit 
\end{deluxetable}

As we have shown in \S~3.3, the power law model yields the best fit to
the \feii\ resonance line data. Similar results are obtained for the
five \hei\ lines that arise from the meta-stable level 2$^3$S, as well
as for the low-excitation levels of the \feii\ ground state.  In
addition to the five resonance lines lines (i.e., E=0) we studied so
far UV Multiplets 1,2 and 3 of \feii\ contain 31 lines from four low
excitation levels (with energies between 385-977 cm$^{-1}$; see table
1 of de~Kool et~al.\ 2001). Troughs from these lines will provide the
diagnostics for finding the number density of the outflow and its
ionization equilibrium (Korista et al.\ 2008). In the contest of this
paper, these lines supply further constraints regarding, which of our
two main models (partial covering or power-law) better represent the
distribution of the absorbing material in front of the source.

In order to test which model works better, and to extract the crucial
column densities of these energy levels, we again used $\chi^2$
minimization. Since all these levels arise from the same ion, their
distribution in front of the source should be similar to that of the
resonance lines, the only freedom we allow is the ratio of level
populations as this depends on the density and temperature of the
absorber.  Therefore, our procedure is to use the same $C(v)$ and
$a(v)$ we found for the partial covering or power-law, respectively,
and vary $\tau(v)$ and $ \tau_{\rm{max}}(v)$ in order to find an
optimal solution.  This is a robust and powerful test since the
absorption troughs for each level are a completely independent set of
measurements from those of the resonance lines and from each other.  In
addition, we also observe five lines from the \hi\ metastable level at
159856 cm$^{-1}$.  Although this material does not have to be
distributed identically to the \feii\ distribution, this is still the
simplest and most constraining assumption.  Therefore we use the exact
same methodology on the \hei*\ lines.

The results are shown in table \ref{excited_level_chi2}.
It is evident that for each of the five independent \feii\ 
levels and for the \hei*\ lines, the $\chi^2_{\rm red}$ 
of the power-law model is lower than that of the partial covering model.
This supplies a strong support to the result already obtained for the 
resonance \feii\ trough: a two parameter power law model fit the data
 better than a two parameter Partial covering model.
 As we discuss in \S~4, the power-law distribution is also a better physical
model for the absorber in QSO~2359--1241.  For these reasons we adopt
the power-law results for all the troughs observed in the object.  The
few exceptions are blended lines for which we can only give lower
limits.

\clearpage


\section{DISCUSSION}


Understanding the nature of quasar outflows and their influence on the
nuclear and galactic environments, begins with deciphering of
their geometry on both small and large scales. Since all of the
observational information is embedded within the absorption troughs,
the first step is to determine the distribution of absorbing material
that creates the troughs. Advances in our understanding of this issue
will shed light on the microphysics of the outflows, and more importantly
will yield reliable measurements of the ionic column densities. These measurements
in turn are crucial to determining almost every physical aspect of the
outflows: the ionization equilibrium and abundances, gas density,
distance, mass flux and kinetic luminosity.

Until the late 1990s, all the work published on quasar outflows (e.g.,
Korista et al.\ 1992; Turnshek et al.\ 1996; Hamann 1998) used the
apparent optical depth ($\tau_{ap}$) model  to extract column densities.  This
method works very well for ISM and IGM absorbers and corresponds to a
simple gemetrical interpretation for the absorbing medium: a
homogeneous screen in front of the light source.  The well demonstrated
failure of the $\tau_{ap}$ model when applied to AGN outflows troughs (see
sections 3 and 1) motivate us to look more carefully at the geometry
involved.  Relatively secure is the following picture:  For 
quasars, the size of the continuum source is  $\sim10^{16}$ cm
and the size of the  broad emission line region is $\sim10^{18}$ cm (Kaspi et al.\ 2005).
Along our line of sight, there is material flowing towards us with
radial velocities ranging from a few hundred to more than 30,000 \kms.  This
material is, cosmologically speaking, in the vicinity of the quasar,
within 0.1-10,000 pc.\ from the central source: 0.1 pc.\ in BALQSO 1603+3002
(Arav et al 1999b); 25 pc.\ for NGC 3783, (Gabel et al.\ 2005b); 1000
pc.\ for BALQSO 1044+3656 (de Kool et al.\ 2001); 28 kpc in quasar 3C
191 (Hamann et al.\ 2001). The question we try to address here is: what is the
 distribution of the absorbing  material across the quasar emission region

Most of the work that tried to advance beyond the $\tau_{ap}$ model
has used the pure covering factor method (see \S~1 and \S~3). This method
has three advantages: 1) It lends itself to a simple geometrical
picture where the absorbing material covers only part of the
source. 2) It yields perfect fits to doublet troughs, which comprise
the vast majority of high-resolution data that is needed to advance
beyond the $\tau_{ap}$ model.  However, as mentioned in \S~3.2, the perfect fit
arises mainly from fitting two equations with two unknowns. 3) Changes
in the covering factor as a function of velocity were interpreted as
accelerated motion with a small non-radial component (the
dynamical-geometrical model of Arav et al.\ 1999a).

However, the pure covering factor model has some significant
weaknesses as well.  1) It has been known for a long time that troughs of
different ions show different covering factors at the same velocity
(e.g., BALQSO 1603+3002, Arav et al 1999b). Explaining such behavior
involves more complicated variants of the model, chief among them is
abandoning the notion of homogeneous material distribution over the
covered surface.  We see a strong case of the above problem in
QSO~2359--1241 where the best fit covering factor solution for the
excited levels of \feii\ is different (smaller) than the the covering
factor for the E=0 level. 2) While the issue above can be somewhat
addressed using more complicated variants of the model, the following
issue poses a more serious problem. We now have several cases where
the distance of the outflow from the source was obtained by
determining the number density of the outflow and combing it with the
ionization equilibrium solution (knowledge of the ionization
parameter).  Some outflows, including the one studied here, are found
to be at distances of $\sim1000$ pc from the source, and at the same
time show clear evidence that the $\tau_{ap}$ model is inapplicable to
their troughs (e.g., BALQSO 1044+3656, de Kool et al.\ 2001).  At
distance up to $\sim$10 times the size of the emission source, the
dynamical-geometrical model for the changes in the covering factor as
a function of velocity can work well. However, it is very difficult to
envision a scenario where this model can work when the distance of the
outflow from the emission source is $10^3-10^4$ times the size of the
emission source.  We note that the covering factor model does not
suffer from these difficulties for outflows at much closer distances
to the nucleus (e.g., BALQSO 1603+3002 and NGC~3783, mentioned above).

Two main efforts were undertaken to address this serious flaw.  One
effort concentrated on finding a way to interpret the number density
using a model that reduces the inferred distance by 2-3 orders of
magnitude (Everett et al 2002).  This was done by invoking a shielded,
multiphase gas that is made of a continuous low-density wind with
embedded high-density clouds.  The model also requires that the clouds
are dusty and have strong differential dust depletion of iron compared
to magnesium.  While the Everett et al (2002) model solves the
covering factor problem by reducing the inferred distance
substantially, this achievement comes at the heavy price of invoking
several special conditions (an ionization shield, multiphase gas, and
dusty clouds with strong differential dust depletion).

The second effort studied inhomogeneous distribution of absorbing
material across the emission source as an alternative to the pure
partial covering model.  The strong feature of these models is that
they do not require a finely tuned covering factor at large distances
and at the same time do not invoke special physical conditions to do
so. Before we discuss the physical picture of these models we need to
address the history of these models, which at first glance looks
contradictory.  de Kool et al (2002, dKKA), showed that such models (similar
to the ones used here) can yield adequate fits for two quasar
outflows.  In contrast, Arav et al (2005), found that the same models
could not fit the outflow troughs of the Seyfert galaxy Mrk~279.
However, in the Mrk~279 data analyzed by Arav et al (2005), more than
half the flux at the spectral location of the outflow troughs is
contributed by the broad emission line (BEL) of the same transition.
Gabel et al (2005a) determined that for Mrk~279 the outflow covers the
entire continuum source but only a portion of the BEL. This picture
looks physically plausible if the distance of the outflow is
co-spatial or just outside the BEL region. Supporting evidence to this
distance scale comes from the fact that the troughs in Mrk~279 vary on
timescale of $\sim1$ year.  Arav et al (2005) tried to find a simple
inhomogeneous power-law model that will fit the data without
differentiating between the BEL and continuum flux distributions, and
concluded that this is not feasible.  It is not known if more
complicated inhomogeneous models that will differentiate between the
BEL and continuum flux sources could give a good fit to the Mrk~279
data.  In contrast, for QSO~2359--1241 the contribution of the BELs at
the spectral location of the \feii\ troughs is less than 10\%, and the
distance scale is $\sim1000$ pc. (Based on the inferred number density
from \feii* troughs combined with a determination of the ionization
parameter, Arav et al 2008). The combination of these conditions
permits us to ignore the contribution of the separate BELs and yields a
better fit for the power-law model than the pure covering factor model does.

Finally, we need to address the fundamental question: What is the
physical distribution of absorbing material in quasar outflows? In
this work we have showed that a power-law distribution of material is
a better model both statistically and physically than the pure
covering factor model. A power-law model is of course a highly
simplified version of what the real distribution might be. A hint
about the actual situation can be inferred from resolved images of
other astrophysical outflows. Planetary nebula, super nova remnants
and other resolved outflows, tend to show a fragmented or web-like
structure of the emitting material. We speculate that quasar outflows
show similar structures on scales smaller than the size of the
emission source. This is a straight forward way to explain the break
down of the homogeneous absorption screen hypothesis. It is our hope
that future dynamical models are able to yield physically-motivated
predictions with regards to the distribution of outflowing absorbing
material. These predictions could then be tested against the QSO
QSO~2359--1241 and similar data sets.

\section{SUMMARY}

\begin{enumerate}

\item We presented 6.3 hours of VLT/UVES high-resolution (R$\sim$40,000) 
spectroscopic observations of QSO~2359--1241 and identified all the 
absorption features associated with the outflow emanating from this object.

\item The unprecedented high signal-to-noise data from five unblended
troughs of \feii\ resonance lines yielded tight constraints on outflow
trough formation models.

\item We find that power-law distribution models for absorption material
 in front of the emission source gives a better fit to the \feii\ data
 than covering factor models.

\item This finding alleviates the problem of obtaining a velocity dependent
partial covering factor at distances $10^3-10^4$ times larger than the
size of the emission source.


\end{enumerate}

\section*{ACKNOWLEDGMENTS}
We thank  the referee for numerous
valuable suggestions.  We acknowledge support from 
NSF grant number AST 0507772 and from 
NASA LTSA grant NAG5-12867.


\end{document}